\documentclass[aps,preprint]{revtex4}%
\usepackage{amsfonts}
\usepackage{amsmath}
\usepackage{amssymb}
\usepackage{graphicx}%
\providecommand{\U}[1]{\protect\rule{.1in}{.1in}}

\begin{document}
\date{\today}
\title{Cooperating or Fighting with Control Noise \\in the Optimal Manipulation of Quantum Dynamics}
\author{Feng Shuang}
\author{Herschel Rabitz}
\affiliation{Department of Chemistry, Princeton University, Princeton, New Jersey 08544}

\begin{abstract}
This paper investigates the impact of control field noise on the optimal
manipulation of quantum dynamics. Simulations are performed on several
multilevel quantum systems with the goal of population transfer in the
presence of significant control noise. The noise enters as run-to-run
variations in the control amplitude and phase with the observation being an
ensemble average over many runs as is commonly done in the laboratory. A
genetic algorithm with an improved elitism operator is used to find the
optimal field that either fights against or cooperates with control field
noise. When seeking a high control yield it is possible to find fields that
successfully fight with the noise while attaining good quality stable results.
When seeking modest control yields, fields can be found which are optimally
shaped to cooperate with the noise and thereby drive the dynamics more
efficiently. In general, noise reduces the coherence of the dynamics, but the
results indicate that population transfer objectives can be met by
appropriately either fighting or cooperating with noise, even when it is intense.

\end{abstract}

\maketitle

\section{Introduction}

Control of quantum processes is actively being pursued theoretically
\cite{Rice00,Rabitz0364} and experimentally\cite{Walmsley0343,Brixner011}. In
practice control field noise and environmental interactions inevitably will be
present, and the general expectation is that their involvement will be
deleterious towards achieving control\cite{Wusheng036751}. The present paper
only considers the influence of field noise upon the controlled dynamics with
the noise described by shot-to-shot pulse variations, as in the signal average
experiments. A separate work will explore analogous issue for environmental
interactions. Recent studies have considered several aspects of the influence
of laser
noise\cite{Walser92468,JM0010841,Rabitz0263405,Ignacio049009,Akramine984892,Akramine984349}%
, and this work aims to further explore the issue. Operating under closed
loop\cite{Judson921500} in the laboratory will naturally deal with the noise
as best as possible. Using simple models and simulations, this paper will
explore the nature of robustness to noise and the possibility of cooperating
with the noise rather than fighting against it, especially in the strong noise
limit. Such a prospect has foundation in analogous classical stochastic
resonance phenomena\cite{Gammaitoni98223}.

The presence of field noise in quantum control is generally viewed as
problematic, but recent successful experiments at least support the point that
some noise can be
tolerated\cite{Zeidler0123420,Kunde00924,Brixner0157,Levis01709,Daniel03536,Gerber98919}
and that it may have a constructive effect in quantum
dynamics\cite{Grossmann93229,Blanchard94749,Shao9781,Klappauf981203}. A recent
theoretical analysis\cite{Rabitz0263405} on the impact of field noise upon
optimal control indicated that an inherent degree of robustness can be
anticipated by virtue of controlled observable expectation values being
bilinear in the evolution operator and its adjoint. Some simulations of closed
loop experiments also show that robust control is possible, and that properly
designed control fields can fight against
noise\cite{JM0010841,Gross924557,Toth943715}. In some special circumstances,
field noise even may be
helpful\cite{Shuang007192,Vugmeister972522,Mark972516,Pechukas942772,Mantegna96563,Mantegna003025}%
. The presence of field noise may also improve the convergence rate of the
laboratory learning control processes\cite{Gross924557,Toth943715}.

In this paper, we will investigate four model illustrations of optimal control
in the presence of field noise, especially for cases of strong noise. It is
naturally found that field noise is deleterious to achieving control if a high
yield is desired, however we also find robust high yield solutions that are
extremely stable with respect to noise. If a low yield is acceptable, then it
is shown that field noise may even be helpful and the control field can
cooperate with it. The present work primarily aims to demonstrate these
findings in several model illustrations. A follow on paper will analyze the
mathematical and physical origin of the observed behavior\cite{Shuang2004_2}.

Section II will present the model and control formulation used in these
studies. Section III describes the genetic algorithm employed to search for an
optimal control field in the presence of noise, and some special algorithmic
modifications are made for this purpose. Section IV presents a set of control
simulations under various system and noise conditions. Some brief concluding
remarks are given in section V.

\section{The Model System}

The effect of field noise on controlled quantum dynamics will be explored in
the context of population transfer in multilevel systems characterized by the
Hamiltonian $H$,%
\begin{subequations}%
\begin{align}
H  &  =H_{0}-uE(t)\label{Ht}\\
H_{0}  &  =\sum_{v}\varepsilon_{\upsilon}\left\vert \upsilon\right\rangle
\left\langle \upsilon\right\vert \label{H0}%
\end{align}%
\end{subequations}%
where $\left\vert \upsilon\right\rangle $ is an eigenstate of $H_{0}$ and
$\varepsilon_{\upsilon}$ is the associated field-free eigen-energy. The
noise-free control field $E(t)$ is taken to have the form%
\begin{subequations}%
\begin{align}
E_{0}(t)  &  =S(t)\sum_{l}A_{l}^{0}\cos\left(  \omega_{l}t+\theta_{l}%
^{0}\right)  \text{,}\label{E0}\\
S(t)  &  =\exp\left[  -\left(  t-T/2\right)  ^{2}/2\sigma^{2}\right]  \text{,}
\label{S}%
\end{align}%
\end{subequations}%
where $\left\{  \omega_{l}\right\}  $ are the allowed resonant transition
frequencies of the system. The controls are the amplitudes $\left\{  A_{l}%
^{0}\right\}  $ and phases $\left\{  \theta_{l}^{0}\right\}  $.

In each simulation, a reference noise-free optimal control calculation is
first performed with the cost function:%
\begin{subequations}%
%

\begin{align}
J_{0}\left[  E_{0}(t)\right]   &  =\left\vert O\left[  E_{0}(t)\right]
-O_{T}\right\vert ^{2}+\alpha F_{0}\label{J0}\\
F_{0}  &  =\sum_{l}\left(  A_{l}^{0}\right)  ^{2}\text{,} \label{F0}%
\end{align}%
\end{subequations}%
where $O_{T}$ is the target value (expressed as a percent yield in the
applications later) and $O\left[  E_{0}(t)\right]  $ is the outcome produced
by the field $E_{0}(t)$, and $F_{0}$ is the fluence of the control field whose
contribution is weighted by the constant, $\alpha>0$. In the present work,
$O=\left\vert \Psi_{f}\right\rangle \left\langle \Psi_{f}\right\vert $ is a
projection operator for the population of the target state $\left\vert
\Psi_{f}\right\rangle $.

Noise in the laboratory could take on various forms and arise from a number of
sources\cite{Ignacio049009}. In keeping with laboratory practice, the achieved
control will be measured as an ensemble average over the outcome of many
control fields. Here the noise is modeled as run-to-run uncertainties in the
amplitudes and phases in Eq.(\ref{E0})\cite{JM0010841}. This work will take an
extreme conservative view to explore the nature of achievable control when
such noise is large and independent of each other in the different runs. The
noise is simulated by introducing the parameters $\gamma_{A_{l}}$ and
$\gamma_{\theta_{l}}$ which are chosen randomly over the range $\left[
-\Gamma_{A},\Gamma_{A}\right]  $ and $\left[  -\Gamma_{\theta},\Gamma_{\theta
}\right]  $, respectively,
\begin{equation}
A_{l}^{\left(  i\right)  }=A_{l}^{0}+\gamma_{A_{l}}^{\left(  i\right)
}\text{, \ \ \ }\theta_{l}^{\left(  i\right)  }=\theta_{l}^{0}+\gamma
_{\theta_{l}}^{\left(  i\right)  }\text{.} \label{noiseL}%
\end{equation}
What is referred to as "amplitude noise" alone actually allows for phase
changes over the range $\left[  -\Gamma_{A},\Gamma_{A}\right]  $. By exploring
the case of $\Gamma_{\theta}\neq0$, $\Gamma_{A}=0$, the phase contribution
will become clear. Different random values are selected for the amplitudes and
phases over their respective ranges in replicate simulations $i=1,2,\cdots,M$
corresponding to an ensemble of $M$ control experiments. The $i$-th experiment
is driven by field
\begin{equation}
E^{\left(  i\right)  }(t)=S(t)\sum_{l}A_{l}^{\left(  i\right)  }\cos\left(
\omega_{l}t+\theta_{l}^{\left(  i\right)  }\right)  \text{.} \label{Ei}%
\end{equation}
and the net outcome of all the control experiments is the average,%
\begin{equation}
\left\langle O\left[  E^{\left(  i\right)  }(t)\right]  \right\rangle
_{N}=\frac{1}{M}\sum_{i=1}^{M}\left\vert \left\langle \psi\left(  \left[
E^{\left(  i\right)  }(t)\right]  ,T\right)  |\psi_{f}\right\rangle
\right\vert ^{2}\text{,} \label{On}%
\end{equation}
where $\left\vert \psi\left(  \left[  E^{\left(  i\right)  }(t)\right]
,T\right)  \right\rangle $ is the final state driven with the $i$-th field in
Eq.(\ref{Ei}). The bracket notation $\left[  E^{\left(  i\right)  }\left(
t\right)  \right]  $ indicates a functional dependence on the field over the
internal $0\leq t\leq T$. The objective function in the presence of noise is:%

\begin{equation}
J_{N}\left(  A_{l}^{0},\theta_{l}^{0}\right)  =\left|  \left\langle O\left[
E^{\left(  i\right)  }(t)\right]  \right\rangle _{N}-O_{T}\right|  ^{2}+\alpha
F_{0}\text{.} \label{JN}%
\end{equation}

In order to find a robust control field, for which the outcome is optimally
insensitive to the noise, we incorporate the measurement standard deviation
$\sigma$ into the cost function to finally produce:%

\begin{equation}
J\left(  A_{l}^{0},\theta_{l}^{0}\right)  =J_{N}\left(  1+\beta\sigma_{N}
\left[  E^{\left(  i\right)  }(t)\right]  \right)  \text{,} \label{JA}%
\end{equation}
where
\begin{equation}
\sigma_{N}\left[  E^{\left(  i\right)  }(t)\right]  =\sqrt{\left\langle
O^{2}\left[  E^{\left(  i\right)  }(t)\right]  \right\rangle _{N}-\left\langle
O\left[  E^{\left(  i\right)  }(t)\right]  \right\rangle _{N}^{2}} \label{Var}%
\end{equation}
and $\left\langle O^{2}\right\rangle _{N}$ is defined similar to
Eq.(\ref{On}). The coefficient $\beta>0$ balances the magnitudes of the terms.
Since $J$ is minimized, pulses that produce a small variance in the objective
are favored.

\section{A Genetic Algorithm with an improved elitism operator in the presence
of control noise}

Current quantum control experiments\cite{Judson921500} employ genetic
(GA)\cite{Goldberg97} or evolutionary algorithms\cite{Zeidler0123420} to
identify optimal control fields. Simulations have shown
\cite{Gross924557,Toth943715} that GAs can tolerate a high level of noise
because they don't make sharp decisions, allowing for decision errors to be
averaged out over the generations of the learning process. GAs can be very
effective, but they also can be expensive to employ, especially when
optimizing a shot-to-shot stochastic process because a large number of
experiments are needed to perform the ensemble average. Many of the
simulations in this work include significant field noise, and it was found
that at least $M=10^{4}$ runs are needed to get accurate ensemble averages. In
order to speed up the simulations, many strategies have been
proposed\cite{Aizawa9348,Miller97,Seijas02617}. Here we used a simple
approach: only perform full statistical averaging with the most effective
field at each generation, while for intermediate assessments of all other
individuals use a small number of ensemble average runs. The detailed
procedure to select an optimal field consists of the following steps:

1. Initialize the GA. We assume there are two special individuals, one is
denoted as ''\textit{best}'', which is the best member of population; another
is denoted as ''\textit{elitist}'', which will be updated in each generation.
\textit{elitist} is the same as \textit{best} when the GA starts.

2. Perform the standard operations of reproduction, crossover and mutation to
generate a new population for the next generation. Evaluate the fitness of
each individual in the population with a small size ensemble ($M\leq$ $10^{3}$
in this work).

3. Compare the objective values arising from \textit{elitist} and \textit{best
}(the best member of the new generation):

if $J(\mathit{elitist})\leq J(\mathit{best})$ then re-evaluate the cost
function of \textit{elitist} with a large ensemble ($M\geq10^{4}$ in this work);

if $J(\mathit{elitist})>\ J(\mathit{best})$ then re-evaluate the cost function
of \textit{best} with a large ensemble ($M\geq10^{4}$ in this work);

4. Compare the updated cost function value of \textit{elitist} and
\textit{best} from the last step;

if $J(\mathit{elitist})\leq J(\mathit{best})$ then replace the worst member of
population with \textit{elitist}

if $J(\mathit{elitist})>J(\mathit{best})$ then set \textit{elitist} =
\textit{best};

5. Go to step 2, and repeat the GA operations until convergence is fulfilled.

Although there is always a chance that a good field is overlooked by this
procedure, it proved to be effective in finding robust fields. The procedure
is very efficient as it only performs extensive statistical averaging on the
likely most effective fields.

\section{Numerical Simulations}

To demonstrate how quantum optimal control can either fight or cooperate with
field noise, we performed four simulations with simple model systems. The
first three simulations use the single path system in Figure 1(a) while the
last simulation uses the double-path system in Figure 1(b). Equation
(\ref{JA}) is the cost function used to the guide GA choice of the control
field. The average yields shown in all the tables typically involve an
ensemble of $M=10^{4}$ samples. The simulations used the Toolkit
technique\cite{Yip038168} to efficiently solve the Schr\"{o}dinger equation.

\subsection{Model 1}

This model used the single path 5-level system in Figure 1(a) with eigenstates
$\left\vert i\right\rangle $, $i=0,\cdots,4$ of the field free Hamiltonian
$H_{0}$, having only nearest neighbor transitions with the frequencies
$\omega_{01}=1.511$, $\omega_{12}=1.181$, $\omega_{23}=0.761$, $\omega
_{34}=0.553$ in rad$\cdot$fs$^{-1}$, and associated transition dipole moments
$\mu_{01}=0.5855$, $\mu_{12}=0.7079$, $\mu_{23}=0.8352$, $\mu_{34}=0.9281$ in
$1.0\times10^{-30}$ C$\cdot$m. The target time is $T=200$ fs, and the pulse
width in Eq.(\ref{Ei}) is $s=30$ fs. The control objective is to transfer the
population from the initially prepared ground state $\left\vert 0\right\rangle
$ to the final state $\left\vert 4\right\rangle $, such that $O=\left\vert
4\right\rangle \left\langle 4\right\vert $. As a reference control case, the
algorithm is run first without any field noise using the cost function in
Eq.(\ref{J0}). Figure 2 shows the optimal field for the two separate goals
$O_{T}=100\%$ and $2.25\%$. These two extreme cases of high and low target
yield are chosen to illustrate their distinct control behavior in the present
of noise. The choice of the low yield target of $2.25\%$ is arbitrary with
similar behavior found for yields up to $\sim10\%$. In some applications even
making a modest yield of a particular product or state for study would be
significant. The physical criteria in the low yield limit is that sufficient
noise be present to produce at least a small population in the target state.
At the target time, the noise free fields in Figures 2(a1) and 2(b1) drive
$98.7\%$ and $2.2\%$ of the population to target state, respectively.

In order to calibrate the strength of the noise, Table I shows the state
population distributions when the system is only driven by noise of various
amplitudes. When $\Gamma_{A}>0.07$, the influence of the field noise is very
large as it drives more than 40\% population out of the ground state. To
examine the influence of noise when seeking an optimal control field, first
consider the case of aiming for a high yield (i.e., perfect control with the
target yield set at $O_{T}=100\%$ in Eq.(\ref{JN})). Table II shows how the
optimal control fields fight with different types of noise in this case. In
order to reveal the separate contributions of noise upon the optimally
controlled dynamics, Table II shows the yield from noise alone $\left\langle
O\left[  E_{n}^{\left(  i\right)  }(t)\right]  \right\rangle _{N}$ without the
control field being present. Here $E_{n}^{\left(  i\right)  }(t)$ is the
$i$-th member of a pure field noise ensemble:%

\begin{equation}
E_{n}^{\left(  i\right)  }(t)=S(t)\sum_{l}\gamma_{A_{l}}^{(i)}\cos\left(
\omega_{l}t+\gamma_{\theta_{l}}^{\left(  i\right)  }\right)  \text{.}
\label{Eni}%
\end{equation}
The control field is determined optimally in the presence of noise; as a
reference, the yield $O\left[  E_{0}(t)\right]  $ of each optimal field
without noise is presented. The second and third rows of Table II show that
the optimal control can fight against amplitude noise by increasing the
fluence $F_{0}$, although the effect of noise is always deleterious. The
variance $\sigma_{N}$ naturally increases as the amplitude $\Gamma_{A}$ rises.
The last three rows of Table II reveal that optimal control fields can be
found that are fully robust to strong phase noise with small yield variance.
This result may appear to be surprising, as proper adjustment of control field
phase is often viewed as necessary for attaining a good quality control
yield\cite{Shuang2004_2}. Table III further examines this issue where it is
found that robustness at any level of noise assures robustness at other
levels. However, this result alone does not reveal the degree of average
coherence maintained in the dynamics. The shot-to-shot signal averaged density
matrix of the quantum system can be defined as:%
\begin{subequations}%
%

\begin{align}
\rho(t)  &  =\frac{1}{M}\sum_{i=1}^{M}\rho^{\left(  i\right)  }(t)\label{rho}%
\\
\rho^{\left(  i\right)  }(t)  &  =\left\vert \psi\left(  \left[  E^{\left(
i\right)  }(t)\right]  ,t\right)  \right\rangle \left\langle \psi\left(
\left[  E^{\left(  i\right)  }(t)\right]  ,t\right)  \right\vert \label{rhoi}%
\end{align}%
\end{subequations}%
Here, $\left\vert \psi\left(  \left[  E^{\left(  i\right)  }(t)\right]
,t\right)  \right\rangle $ is the state driven by the $i$-th field in
Eq.(\ref{Ei}) from the ensemble. The density matrix is further decomposed into
its coherent and diagonal portions, $R_{c}(t)$, $R_{d}(t)$, respectively,%
\begin{subequations}%
\label{R}%
%

\begin{align}
R_{c}(t)  &  =\sum_{k\neq j}\left\vert \rho_{kj}(t)\right\vert ^{2}\text{;
\ \ \ \ \ }R_{d}(t)=\sum_{k}\left\vert \rho_{kk}(t)\right\vert ^{2}%
\label{Rcd}\\
R(t)  &  =\sum_{k,j}\left\vert \rho_{kj}(t)\right\vert ^{2}=R_{c}%
(t)+R_{d}(t)=\text{Tr}\left[  \rho^{2}(t)\right]  \text{.} \label{Rt}%
\end{align}%
\end{subequations}%
Figure 3 plots these measures of the density matrix and the state populations
of the system driven by the field in the last row of Table II with strong
phase noise $\Gamma_{\theta}=\pi$. Figure 3(a1) shows that the high level of
phase noise totally destroys the coherence of the system ($R_{c}(t)\simeq0$)
while Figure 3(b1) shows that large coherence is built up by the same control
if no noise is present. The state populations are also shown in Figures 3(a2)
and 3(b2) indicating in both cases that the target goal is reached via a
simple four-step ladder climbing pathway which is evidently not sensitive to
phase. The result in Figure 3 needs to be understood in the context that
$\rho^{\left(  i\right)  }(t)$, for the arbitrary $i-th$ member of the
ensemble, will generally have a significant degree of coherence along its
evolution. The details of the shot-to-shot coherence changes rather randomly
such that $R_{c}(t)\simeq0$, but the noise average $\left\langle \text{Tr}%
\rho^{2}\right\rangle $ could contain a coherent contribution. Further insight
into this situation will be analyzed upon examination of model 2 in Section 4.2.

If we accept a low control yield outcome, very different control behavior is
found in the presence of noise. The results from optimizing Eq.(\ref{JA}) are
shown in Table IV with various levels of amplitude and phase noise. In all
cases the target yield is $O_{T}=2.25\%$. Table IV shows that low field
amplitude noise ($\Gamma_{A}\leq0.01$) has little impact on the control
outcome. Additional amplitude noise becomes more helpful indicated by the
reduced control field fluence. In the domain $0.06\leq\Gamma_{A}$ $\leq0.08$,
the control outcome is much larger than the sum of the yield from the control
field and noise alone. For example at $\Gamma_{A}=0.07$ the noise and optimal
field alone, respectively, produce yields of $0.26\%$ and $0.16\%$. But, the
same field operating in the presence of the noise produces a yield $2.11\%$.
This behavior indicates that the field is cooperating with the presence of
noise to more effectively achieve the posed goal. The left and right variance,
$\sigma_{N}^{-}$ and $\sigma_{N}^{+}$, respectively, indicate a very
asymmetric distribution having a long tail on the high population side. Again,
we also find several solutions that are very stable to phase noise. The
collective results for amplitude noise are plotted in Figure 4.

Figure 5 shows the evolution of the density matrix measures $R_{c}$, $R_{d}$
and $R$ in Eq.(\ref{R}) at different field noise amplitude $\Gamma_{A}$
levels. With strong noise there is little surviving coherence indicated by
$R_{c}$, except in the time around t$\sim100$ fs. In order to further explore
the robust nature of the controlled dynamics in this low yield regime, Table V
presents the outcome of applying noise at the amplitude $\Gamma_{A}^{\prime}$
to a field optimally determined at noise amplitude $\Gamma_{A}$. The yield for
$E_{\Gamma_{A}}^{op}$ with $\Gamma_{A}^{\prime}=\Gamma_{A}$ is the outcome for
the field optimally determined with the noise at $\Gamma_{A}$. It is evident
that once a field cooperates with noise at level $\Gamma_{A}$, then it
cooperates with noise at any other level. We may conclude that at low yield,
cooperation can be arranged, but at the expense of coherence.

\subsection{Model 2}

Model 2 extends the Hamiltonian of model 1 in Figure 1(a) to allow for two
quanta transitions: $\omega_{02}=2.692$, $\omega_{13}=1.942$, $\omega
_{24}=1.314$ in rad$\cdot$fs$^{-1}$, with transition dipole elements:
$\mu_{02}=-0.1079$, $\mu_{13}=-0.1823$, $\mu_{24}=-0.2786$, in $1.0\times
10^{-30}$ C$\cdot$m. Table VI shows that for a high target yield of
$O_{T}=100\%$, the control field can again effectively fight against amplitude
noise by increasing the fluence $F_{0}$. But, as with the results in Table II,
increasing the amplitude of the noise makes this battle harder with the final
average yield decreasing and the variance increasing. The case of pure phase
noise at $\Gamma_{\theta}=1.0$ shows sensitivity to its presence with the
yield falling to $68\%$. The last row of Table VI also shows a solution that
is extremely robust to phase noise with reduced variance at a surprisingly
small fluence(i.e., the fluence is even smaller than for the case $\Gamma
_{A}=0.0$.) This latter behavior is likely the result of there being many
solutions to the optimal control search process with the algorithm finding
alternative solutions under different circumstances.

Table VII further analyzes the robustness of the last two phase noise
contaminated fields in Table VI. When the field $E_{0.0}^{op}$ (i.e.,
determined with no noise) is applied with a high level of phase noise, at
$\Gamma_{\theta}=2.0$, there is poor robustness with the final yield dropping
to 24.2\%. However, for the yield optimally determined at the high noise level
$\Gamma_{\theta}^{\prime}=2.0$, there is robustness to lower levels of noise.
The power spectra for the field optimized at $\Gamma_{\theta}=0.0$ and $2.0$
are shown in Figure 6. Comparing the two plots we see that $E_{2.0}^{op}(t)$
drives the system along a path of single quantum transitions while
$E_{0.0}^{op}(t)$ drives the system as well with the additional two quanta
transition $2\rightarrow4$. The presence of the extra two quantum transition
can set up a delicate situation which is sensitive to phase noise, as Table
VII indicates. The robustness of the single quantum path for $\Gamma_{\theta
}=2.0$ is analogous to what was found in model 1.

Two tables similar as Table IV and V but for model 2 can be constructed to
show strong cooperation between optimal control field and noise when target
yield is low ($O_{T}=3.65\%$), and the data can be plotted to form a figure
analogous to that of Figure 4.

\subsection{Model 3}

This model is same as model 1, but amplitude noise is only present in the
$0\leftrightarrow1$ transition. Thus, noise alone cannot drive any population
into the target state. There is also no phase noise in any of the transitions.
Table VIII shows the results from different levels of $0\leftrightarrow1$
amplitude noise for a low yield target of $O_{T}=10\%$. There is little
evidence of cooperation with the noise, except in the last row with
$\Gamma_{A}=0.2$. In this case, the field fluence is low and the yield from
the field alone is almost zero. But, the desired target is achieved when the
field cooperates with the noise. Figure 7 shows the power spectrum of the
optimal field for this case and it is evident that the $0\leftrightarrow1$
transition is absent. The mechanism of cooperation is clear: noise drives the
system from level $0$ to level $1$, then the control field drives it to level
$4$. The remaining cases in Table VIII do not exclude the possibility that
fields could be found that also cooperate with other levels of noise. The
result at $\Gamma_{A}=0.20$ shows that such noise cooperative solutions exist.

\subsection{Model 4}

The fourth model is the more complex two-path system in Figure 1(b). In this
model, population can be transferred to the target state $\left\vert
4\right\rangle $ along two separate pathways. The transition frequencies and
dipoles of the left path are same as that of model 1; the right path has the
distinct transition frequencies: $\omega_{0^{\prime}1^{\prime}}=2.153$,
$\omega_{1^{\prime}2^{\prime}}=1.346$, $\omega_{2^{\prime}3^{\prime}}=0.345$,
$\omega_{3^{\prime}4^{\prime}}=0.162$ in rad$\cdot$fs$^{-1}$ and dipole
elements: $\mu_{0^{\prime}1^{\prime}}=0.6526$, $\mu_{1^{\prime}2^{\prime}%
}=0.7848$, $\mu_{2^{\prime}3^{\prime}}=0.9023$, $\mu_{3^{\prime}4^{\prime}%
}=1.0322$ in $1.0\times10^{-30}$ C$\cdot$m. For simplicity, along both paths
only single quanta transitions are allowed.

The results in Table IX show cooperation between the field and the noise in
model 4 for a low target yield of $O_{T}=2.25\%$. Figure 8 shows the power
spectra of the three optimal fields in Table IX. Panel (a) indicates that in
the case of no noise the field primarily drives system along the right path.
When noise is present in the left path (panel (b)), the optimal field drives
the system along the left path in order to cooperate with the noise.
Similarly, when the noise is in the right path (panel (c)), the optimal field
drives the system along the right path. An interesting feature of case (c) is
the lack of intensity for the last step $\left\vert 3^{\prime}\right\rangle
\rightarrow\left\vert 4^{\prime}\right\rangle $ along the path. Evidently the
optimal field fully draws on the noise for this step as a special case of
cooperation. The presence of the fluence term in the cost function favors such
behavior provided that the target goal can be met.

These results clearly show that efficiently achieving the yield is the guiding
principal dictating the nature of the control field and mechanism in this
case. This behavior reflects the nature of the cost function in Eq.(\ref{JN}),
which places no demand on the coherence of the dynamics, but only asks that
the control be efficiently achieved with $\alpha>0$. If coherence maintenance
was also included in the cost function the field would be guided accordingly,
in keeping with the ability to meet the newly posed objective.

\section{Conclusions}

The impact of amplitude and phase noise upon quantum control is explored in
this paper with particular emphasis on these being strong. Numerical
simulations from several cases suggest that control fields can be found that
either cooperate with or fight against noise, depending on the circumstances.
In the case of low target yields, the control field can effectively cooperate
with amplitude noise to drive the dynamics while minimizing the control
fluence. In this regime pure phase noise has little impact, if the pathway to
the target involved single quanta transitions. But, phase noise was found to
be disruptive if there are multiple interfering pathways. Optimal solutions
were also found that exhibited stability with respect to phase noise while
maintaining high yields. In general, when successfully either fighting or
cooperating with significant noise, the expense paid is a measured loss of
dynamical coherence. In some applications a lack of full coherence can be tolerated.

The interaction between the noise and field driven dynamics is generally a
highly nonlinear process. The noise can be constructive\cite{Klappauf981203}
or destructive\cite{Ignacio049009} in the manipulation of quantum dynamics,
improve\cite{Gross924557, Toth943715} or reduce\cite{Rana96198} the
convergence rate of searching processes, and this disparate behavior makes it
difficult to precisely identify the impact of noise under various
circumstances. This work explored several special cases to reveal what may
happen when searching for optimal controls in the presence of field noise; a
mathematical analysis of the observed behavior may also be carried
out\cite{Shuang2004_2}. The ability to find robust, and even cooperative
solutions, is most encouraging. These results are consistent with a recent
analysis \cite{Rabitz0263405} arguing that optimally controlled quantum
dynamics has a special inherent degree of robustness. Although, each case will
likely have its own unique characteristic behavior, the findings are
encouraging for the ability to manipulate quantum dynamics even with severe
noise being present. It would also be interesting to explore these issues in
the laboratory by selectively introducing measured levels of noise into the
control field.\bigskip

\begin{acknowledgments}
The author acknowledge support from the National Science Foundation and an ARO
MURI grant. We gratefully acknowledge many valuable conservation with Mark
Dykman on this work.
\end{acknowledgments}

\pagebreak

\ \ \ Table I. Population distribution of the single path model 1 when the
system is only driven by amplitude noise.

\noindent%
\begin{ruledtabular}%
\begin{tabular}
[c]{cccccc}
& \multicolumn{5}{c}{Population in the state}\\\cline{2-6}%
$\Gamma_{A}\;^{a}$ & 0 & 1 & 2 & 3 & 4\\\hline
.10 & 41.6\% & 32.1\% & 17.9\% & 6.1\% & 2.25\%\\
.09 & 44.0\% & 34.4\% & 16.0\% & 4.4\% & 1.24\%\\
.08 & 47.8\% & 35.9\% & 12.9\% & 2.8\% & 0.63\%\\
.07 & 54.1\% & 34.1\% & 9.8\% & 1.7\% & 0.26\%\\
.05 & 69.8\% & 26.0\% & 3.9\% & 0.35\% & 0.026\%\\
.03 & 86.8\% & 12.5\% & 0.68\% & 0.023\% & 0.0006\%\\
.01 & 98.4\% & 1.6\% & 0.0096\% & 3.6$\times10^{-5}$\% & 1.0$\times10^{-7}%
$\%\\
.00 & 100\% & 0\% & 0\% & 0\% & 0\%
\end{tabular}%
\end{ruledtabular}%

$^{a}$ The nominal field is zero: $A_{l}^{0}=0$, and there is no phase noise:
$\gamma_{\theta_{l}}^{\left(  i\right)  }=0.0$; $\Gamma_{A}$ is the maximum
noise amplitude: $\left\vert \gamma_{A_{l}}^{\left(  i\right)  }\right\vert
\leq\Gamma_{A}$, in Eq.(\ref{noiseL}).\pagebreak

Table II. Optimal control field fighting against amplitude and phase noise
with model 1 for the high target yield $O_{T}=100\%$\noindent

\noindent%
\begin{ruledtabular}%
\begin{tabular}
[c]{ccccccc}%
$\Gamma_{A}$,$\Gamma_{\theta}\;^{a}$ & $\left\langle O\left[  E_{n}^{\left(
i\right)  }(t)\right]  \right\rangle _{N}\;^{b}$ & $F_{0}\;^{c}$ & $O\left[
E_{0}(t)\right]  \;^{d}$ & $F_{n}\;^{e}$ & $\left\langle O\left[  E^{\left(
i\right)  }(t)\right]  \right\rangle _{N}\;^{f}$ & $\sigma_{N}$\\\hline
0,0 & 0.00\% & 0.066 & 98.7\% & 0.00 & 98.7\% & 0\%\\
0.05,0 & 0.026\% & 1.73 & 98.6\% & $3.3\times10^{-3}$ & 89.4\% & 5.31\%\\
0.10,0 & 2.25\% & 12.3 & 95.8\% & $1.3\times10^{-2}$ & 73.6\% & 13.1\%\\
0,1 & - & 0.065 & 97.8\% & - & 97.1\% & 0.65\%\\
0,2 & - & 0.065 & 95.9\% & - & 96.2\% & 1.04\%\\
0,$\pi$ & - & 0.065 & 97.7\% & - & 96.0\% & 1.08\%
\end{tabular}%
\end{ruledtabular}%

$^{a}$ $\Gamma_{A}$, $\Gamma_{\theta}$ are the maximum of the amplitude and
phase noise, respectively: $\left|  \gamma_{A_{l}}^{\left(  i\right)
}\right|  \leq\Gamma_{A}$, $\left|  \gamma_{\theta_{l}}^{\left(  i\right)
}\right|  \leq\Gamma_{\theta}$.

$^{b}\;\left\langle O\left[  E_{n}^{\left(  i\right)  }(t)\right]
\right\rangle _{N}$: Yield from noise alone (Eq. (\ref{Eni})) without a
control field.

$^{c}\;$Fluence of the control field (Eq.(\ref{F0})).

$^{d}$ $O\left[  E_{0}(t)\right]  $: Yield arising from the control field
(Eq.(\ref{E0})) without noise.

$^{e}\;$Fluence of the noise.

$^{f}\;\left\langle O\left[  E^{\left(  i\right)  }(t)\right]  \right\rangle
_{N} $: Yield from the optimal control field in the presence of
noise(Eq.(\ref{On}))\pagebreak

Table III. Yield from optimal control fields in the case of model 1 with
different levels of phase noise for the high target yield of $O_{T}=100\%$.

\noindent%
\begin{ruledtabular}%
\begin{tabular}
[c]{ccccc}
& \multicolumn{4}{c}{$\left\langle O\left[  E_{\Gamma_{\theta}}^{op}%
(t)\right]  \right\rangle _{\Gamma_{\theta}^{\prime}}$ $^{a}$}\\\cline{2-5}%
$\Gamma_{\theta}^{\prime}$ & $E_{0.0}^{op}$ & $E_{^{1.0}}^{op}$ &
$E_{2.0}^{op}$ & $E_{\pi}^{op}$\\\hline
0.0 & 98.7\% & 97.8\% & 95.9\% & 97.7\%\\
1.0 & 97.2\% & 97.1\% & 96.4\% & 96.7\%\\
2.0 & 95.4\% & 96.1\% & 96.2\% & 96.1\%\\
$\pi$ & 95.1\% & 95.9\% & 96.1\% & 96.0\%
\end{tabular}%
\end{ruledtabular}%

\bigskip$^{a}$ Yield at the field $E_{\Gamma_{\theta}}^{op}(t)$ with phase
noise: $\left\vert \gamma_{\theta_{l}}^{\left(  i\right)  }\right\vert
\leq\Gamma_{\theta}^{\prime}$. Here $E_{\Gamma_{\theta}}^{op}(t)$ is the field
for which the yield was optimized (i.e., the diagonal elements in the Table)
at the phase noise level $\Gamma_{\theta}$.\pagebreak

Table IV. Optimal control with both amplitude and phase noise for model 1 with
a low objective yield of $O_{T}=2.25\%$.

\noindent%
\begin{ruledtabular}%
\begin{tabular}
[c]{ccccccc}%
$\Gamma_{A}$,$\Gamma_{\theta}{}^{a}$ & $\left\langle O\left[  E_{n}^{\left(
i\right)  }(t)\right]  \right\rangle _{N}^{b}{}$ & $F_{0}\;^{c}$ & $O\left[
E_{0}(t)\right]  ^{d}$ & $F_{n}\;^{e}$ & $\left\langle O\left[  E^{\left(
i\right)  }(t)\right]  \right\rangle _{N}^{f}{}$ & $\sigma_{N}(\sigma_{N}%
^{-},\sigma_{N}^{+})\;^{g}$\\\hline
0.10,0 & 2.25\% & 0.00 & 0.00\% & 1.33$\times10^{-2}$ & 2.25\% &
4.7\%(2.0\%,10\%)\\
0.09,0 & 1.24\% & 3.1$\times10^{-3}$ & 2$\times10^{-6}$ & 1.02$\times10^{-2}$
& 2.09\% & 5.8\%(1.8\%,12\%)\\
0.08,0 & 0.63\% & 6.1$\times10^{-3}$ & 2.5$\times$10$^{-5}$ & 8.53$\times
10^{-3}$ & 2.17\% & 6.5\%(2.0\%,15\%)\\
0.07,0 & 0.26\% & 8.4$\times10^{-3}$ & 0.16\% & 6.53$\times10^{-3}$ & 2.11\% &
5.9\%(2.0\%,13\%)\\
0.05,0 & 0.026\% & 9.9$\times10^{-3}$ & 1.34\% & 3.33$\times10^{-3}$ &
2.08\% & 4.7\%(1.8\%,9.3\%)\\
0.03,0 & 6.0$\times10^{-4}$\% & 1.06$\times10^{-2}$ & 1.87\% & 1.20$\times
10^{-3}$ & 2.06\% & 2.6\%(1.4\%,4.1\%)\\
0.01,0 & 1.0$\times10^{-7}$\% & 1.11$\times10^{-2}$ & 2.12\% & 1.33$\times
10^{-4}$ & 2.17\% & 0.8\%(0.7\%,1.0\%)\\
0,0.5 & \textit{-} & 1.09$\times10^{-2}$ & 2.18\% & - & 2.17\% & 0.01\%\\
0,1.0 & \textit{-} & 1.10$\times10^{-2}$ & 2.21\% & - & 2.18\% & 0.03\%\\
0,2.0 & \textit{-} & 1.10$\times10^{-2}$ & 2.19\% & - & 2.17\% & 0.03\%\\
0,0 & - & 1.09$\times10^{-2}$ & 2.20\% & - & 2.20\% & 0.00\%
\end{tabular}%
\end{ruledtabular}%

$^{a,b,c,d,e,f}$ refer to Table II.

$^{g}\;\sigma_{N}^{-}$, $\sigma_{N}^{+}$ are left and right standard
deviation, respectively, around the mean in $^{f}$.\pagebreak

Table V. Yield from optimal control fields with different levels of amplitude
noise for model 1.

\noindent%
\begin{ruledtabular}%
\begin{tabular}
[c]{ccccccccc}
& \multicolumn{8}{c}{$\left\langle O\left[  E_{\Gamma_{A}}^{op}(t)\right]
\right\rangle _{\Gamma_{A}^{\prime}}$ $^{a}$}\\\cline{2-9}%
$\Gamma_{A}^{\prime}$ & $E_{0.10}^{op}$ & $E_{0.09}^{op}$ & $E_{0.08}^{op}$ &
$E_{0.07}^{op}$ & $E_{0.05}^{op}$ & $E_{0.03}^{op}$ & $E_{0.01}^{op}$ &
$E_{0.00}^{op}$\\\hline
0.10 & 2.25 & 2.98 & 4.15 & 4.35 & 4.64 & 4.74 & 5.25 & 5.50\\
0.09 & 1.24 & 2.09 & 2.96 & 3.80 & 3.99 & 4.11 & 4.79 & 4.53\\
0.08 & 0.63 & 1.27 & 2.17 & 2.92 & 3.28 & 3.60 & 4.16 & 3.76\\
0.07 & 0.26 & 0.71 & 1.48 & 2.11 & 2.90 & 3.18 & 3.48 & 3.46\\
0.05 & 0.026 & 0.17 & 0.55 & 1.1 & 2.08 & 2.50 & 2.91 & 3.36\\
0.03 & 6.0$\times10^{-4}$ & 0.023 & 0.14 & 0.44 & 1.58 & 2.06 & 2.38 & 2.45\\
0.01 & 1.0$\times10^{-7}$ & 0.0014 & 0.016 & 0.19 & 1.38 & 1.88 & 2.17 &
2.21\\
0.00 & 0.00 & 0.00020 & 0.0025 & 0.16 & 1.34 & 1.87 & 2.12 & 2.20
\end{tabular}%
\end{ruledtabular}%

\bigskip$^{a}$ Yield at $E_{\Gamma_{A}^{\prime}}^{op}(t)$ with noise bounded
by $\left\vert \gamma_{A_{l}}\right\vert \leq\Gamma_{A}^{\prime}$ and for
$\gamma_{\theta}=0$. Here $E_{\Gamma_{A}}^{op}(t)$ is field for which the
yield (i.e., the diagonal elements in the table) was optimized with noise
$\Gamma_{A}^{\prime}=\Gamma_{A}$.\pagebreak\ The yields are expressed in percent.

\bigskip Table VI. The yield from optimal control fields fighting with
amplitude or phase noise in model 2 with the high yield target objective of
$O_{T}=100\%$.

\noindent%
\begin{ruledtabular}%
\begin{tabular}
[c]{ccccccc}%
$\Gamma_{A}$,$\Gamma_{\theta}\;^{a}$ & $\left\langle O\left[  E_{n}^{\left(
i\right)  }(t)\right]  \right\rangle _{N}\;^{b}$ & $F_{0}\;^{c}$ & $O\left[
E_{0}(t)\right]  \;^{d}$ & $F_{n}\;^{e}$ & $\left\langle O\left[  E^{\left(
i\right)  }(t)\right]  \right\rangle _{N}\;^{f}$ & $\sigma_{N}$\\\hline
0.0,0.0 & 0.00\% & 0.28 & 99.1\% & 0.00 & 99.1\% & 0.00\%\\
0.05,0.0 & 0.13\% & 1.53 & 98.2\% & 5.83$\times10^{-3}$ & 87.2\% & 7.4\%\\
0.1,0.0 & 3.65\% & 3.27 & 93.7\% & 2.33$\times$10$^{-2}$ & 66.5\% & 14.5\%\\
0.0,1.0 & - & 0.33 & 80.4\% & - & 68.0\% & 13.1\%\\
0.0,2.0 & - & 0.065 & 96.5\% & - & 91.6\% & 4.2\%
\end{tabular}%
\end{ruledtabular}%

$^{a,b,c,d,e,f}$ refer to Table II\pagebreak

Table VII. Control yields with optimal fields at different phase noise levels
for the high yield target $O_{T}=100\%$ in case of model 2

\noindent%
\begin{ruledtabular}%
\begin{tabular}
[c]{cccc}
& \multicolumn{3}{c}{$\left\langle O\left[  E_{\Gamma_{\theta}}^{op}%
(t)\right]  \right\rangle _{\Gamma_{\theta}^{\prime}}$ $^{a}$}\\\cline{2-4}%
$\Gamma_{\theta}^{\prime}$ & $E_{0.0}^{op}$ & $E_{1.0}^{op}$ & $E_{2.0}^{op}%
$\\\hline
0.0 & 99.1\% & 80.4\% & 96.5\%\\
1.0 & 42.9\% & 68.0\% & 95.3\%\\
2.0 & 24.2\% & 37.5\% & 91.6\%
\end{tabular}%
\end{ruledtabular}%

\qquad$^{a}$ refer to Table III\pagebreak

Table VIII. Control outcomes from optimal fields for model 3 with a low target
yield of $O_{T}=10\%$.

\noindent%
\begin{ruledtabular}%
\begin{tabular}
[c]{cccccc}%
$\Gamma_{A}\;^{a}$ & $\left\langle P_{1}\left[  E_{n}^{\left(  i\right)
}(t)\right]  \right\rangle _{N}\;^{b}$ & $F_{0}\;^{c}$ & $O\left[
E_{0}(t)\right]  \;^{d}$ & $\left\langle O\left[  E^{\left(  i\right)
}(t)\right]  \right\rangle _{N}\;^{f}$ & $\sigma_{N}$\\\hline
0.00 & 0.00\% & 0.0176 & 9.95\% & 9.95\% & 0.00\%\\
0.01 & 1.67\% & 0.0198 & 9.91\% & 9.94\% & 1.39\%\\
0.05 & 31.7\% & 0.0229 & 10.66\% & 9.97\% & 3.24\%\\
0.10 & 61.1\% & 0.0194 & 9.27\% & 9.99\% & 8.09\%\\
0.20 & 46.6\% & 0.0144 & 0.0013\% & 9.90\% & 6.06\%
\end{tabular}%
\end{ruledtabular}%

$^{a}\;$Amplitude noise is only in the transitions $0\leftrightarrow1$.

$^{b}\;$Population in level 1 when the system is only driven by noise.

$^{c,d,f}$ refer to Table II.\pagebreak

Table IX. Yield attained from the optimal fields for model 4 with the low
yield target of $O_{T}=2.25\%$.

\noindent%
\begin{ruledtabular}%
\begin{tabular}
[c]{ccccc}%
$\Gamma_{A}^{\left(  L\right)  },\Gamma_{A}^{\left(  R\right)  }\;^{a}$ &
$\left\langle O\left[  E_{n}^{\left(  i\right)  }(t)\right]  \right\rangle
_{N}\;^{b}$ & $F_{0}\;^{c}$ & $O\left[  E_{0}(t)\right]  ^{d}$ & $\left\langle
O\left[  E^{\left(  i\right)  }(t)\right]  \right\rangle _{N}\;^{f}$\\\hline
0.00,0,00 & 0.00\% & 0.0093 & 2.19\% & 2.19\%\\
0.08,0.00 & 0.63\% & 0.0064 & 0.0003\% & 2.17\%\\
0.00,0.07 & 0.50\% & 0.0052 & 0.0002\% & 2.08\%
\end{tabular}%
\end{ruledtabular}%

$^{a}$\ $\Gamma_{A}^{\left(  L\right)  },\Gamma_{A}^{\left(  R\right)  }$:
Noise in the left and right paths, respectively;

$^{b,c,d,f}$ refer to Table II.

\pagebreak

\bigskip%
\begin{figure}
[ptb]
\begin{center}
\includegraphics[
trim=0.000000in 0.000000in -0.853947in 0.000000in,
height=4.1174in,
width=6.1761in
]%
{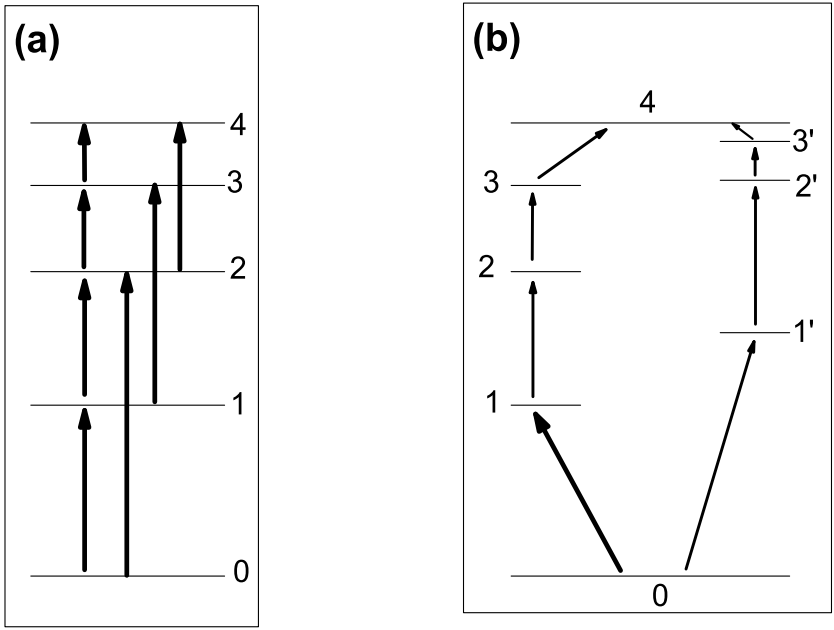}%
\caption{Two quantum multilevel systems used to investigate the impact of
noise on optimally controlled dynamics. (a) The 5 level single path system
used for simulations with models 1,2,3 in Sections 4.1, 4.2, and 4.3,
respectively. Model 1 and 3 only allow single quanta transitions while model 2
allows both single and double quanta transitions. (b) The double-path system
for model 4 in Section 4.4.}%
\end{center}
\end{figure}
\begin{figure}
[ptb]
\begin{center}
\includegraphics[
trim=0.000000in 0.000000in -0.659976in 0.000000in,
height=4.1174in,
width=6.1769in
]%
{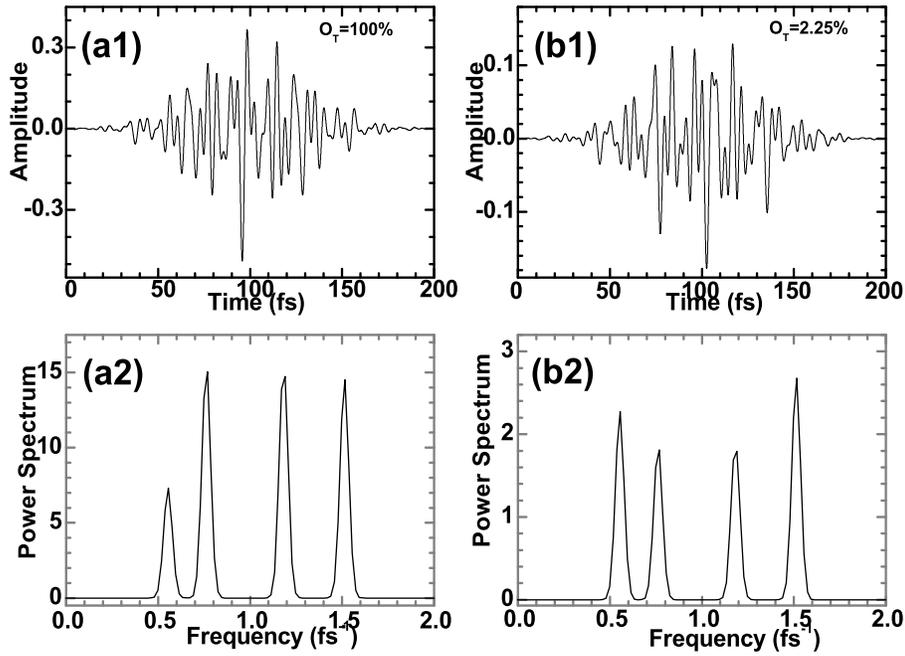}%
\caption{Optimal control fields and their power spectra for model 1 when the
field has no noise. The fields are found using the cost function in
Eq.(\ref{J0}). Plots (a1) and (a2) correspond to the high target yield of
$O_{T}=100\%$ while plots (b1) and (b2) correspond to the low target yield of
$O_{T}=2.25\%$.}%
\end{center}
\end{figure}
\begin{figure}
[ptb]
\begin{center}
\includegraphics[
trim=0.000000in 0.000000in -0.743874in 0.000000in,
height=4.1174in,
width=6.1769in
]%
{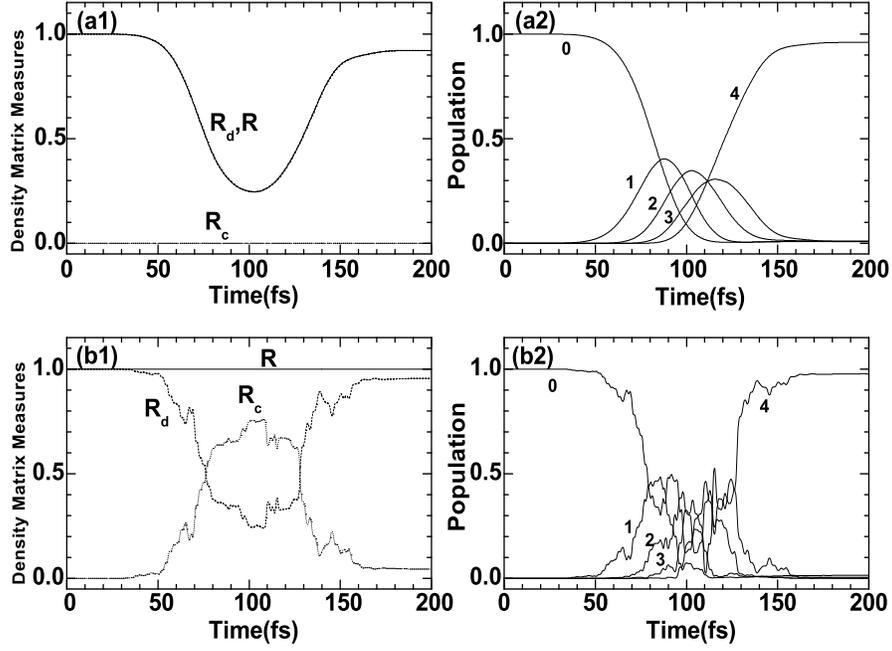}%
\caption{The time dependence of model 1 dynamics driven by the optimal control
field in the last row of Table II. The density matrix measures $R_{d}(t)$,
$R_{c}(t)$, $R(t)$ in plots (a1) and (b1) are defined in Eq.(\ref{R}). Plots
(a1) and (a2) show the dynamics when the system is driven by a control field
with noise while plots (b1) and (b2) show the dynamics of the system driven by
the same field but without noise. The associated state populations are shown
in plots (a2) and (b2).}%
\end{center}
\end{figure}
\begin{figure}
[ptb]
\begin{center}
\includegraphics[
trim=0.000000in 0.000000in -1.230395in 0.000000in,
height=4.1174in,
width=6.1769in
]%
{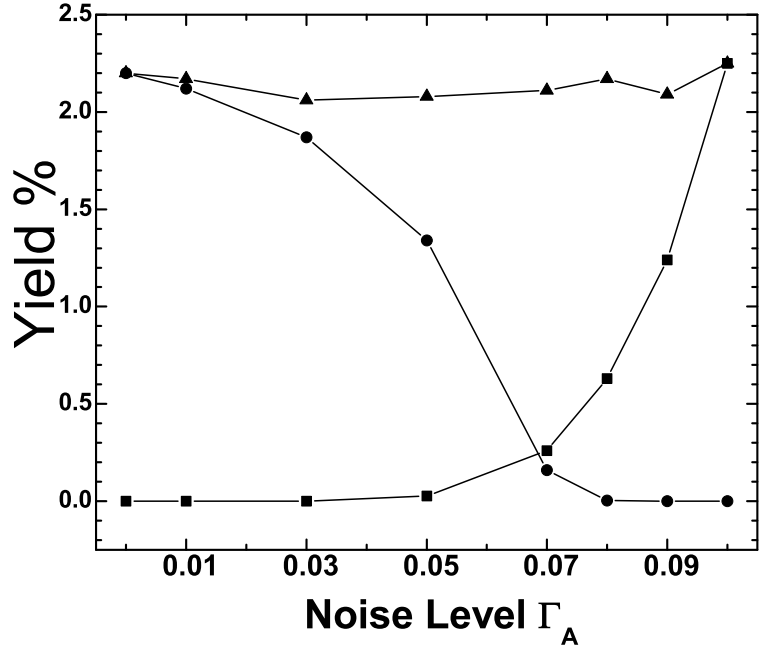}%
\caption{The control yield under various noise conditions for model 1 with the
low yield target of $O_{T}=2.25\%$ in Table V. The labeled curves show the
yield for noise alone, the yield for the optimal field alone and the yield for
the field in the presence of noise. There is notable cooperation between the
noise and the field especially over the amplitude noise range $0.06\leq
\Gamma_{A}\leq0.08$.}%
\end{center}
\end{figure}
\begin{figure}
[ptb]
\begin{center}
\includegraphics[
trim=0.000000in 0.000000in -0.870771in 0.000000in,
height=4.1174in,
width=6.1769in
]%
{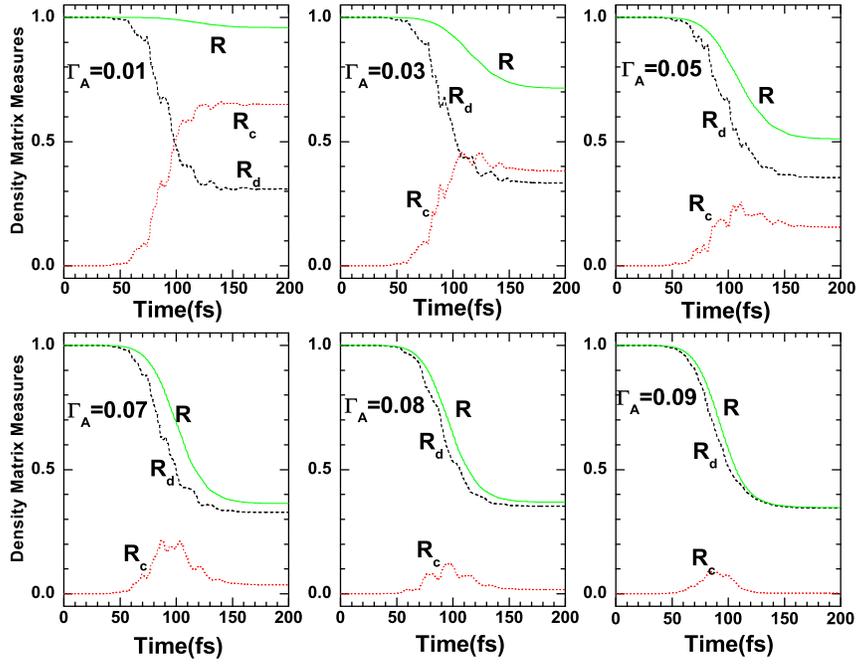}%
\caption{Evolution of the density matrix measures in Eq.(12) with different
levels of field amplitude noise for the low target yield case of Figure 4. The
coherence is successively destroyed by stronger noise, but it still survives
around t$\sim100$ fs.}%
\end{center}
\end{figure}
\begin{figure}
[ptb]
\begin{center}
\includegraphics[
trim=0.000000in 0.000000in -1.037895in 0.000000in,
height=4.1174in,
width=6.1769in
]%
{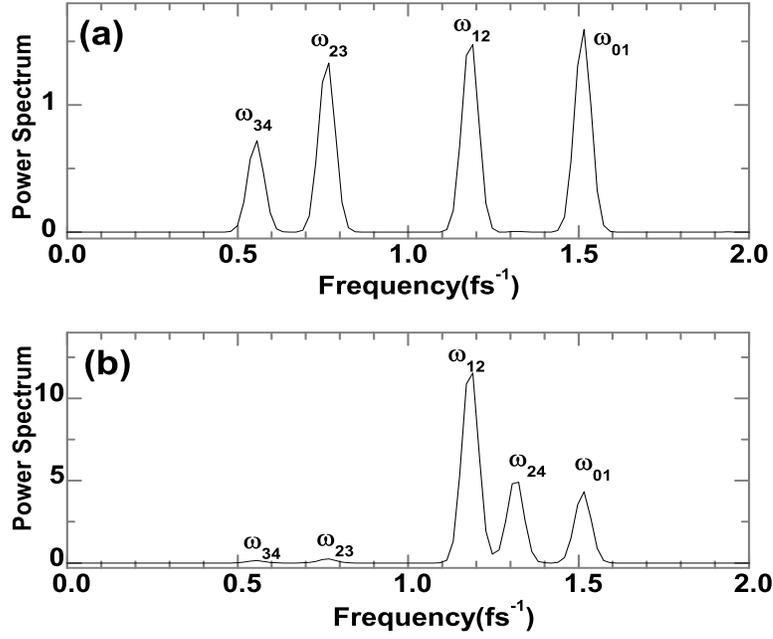}%
\caption{Power spectra of the control fields for model 2 aiming at a high
yield with $O_{T}=100\%$. Plot(a) is for the field $E_{2.0}^{op}$ robust to
phase noise; plot(b) is for the field $E_{0.0}^{op}$ which is not robust to
noise. The robust control field in (a) avoids the extra interference arising
from the two quanta transition $2\leftrightarrow4$.}%
\end{center}
\end{figure}
\begin{figure}
[ptb]
\begin{center}
\includegraphics[
trim=0.000000in 0.000000in -0.256982in 0.000000in,
height=4.1174in,
width=6.1769in
]%
{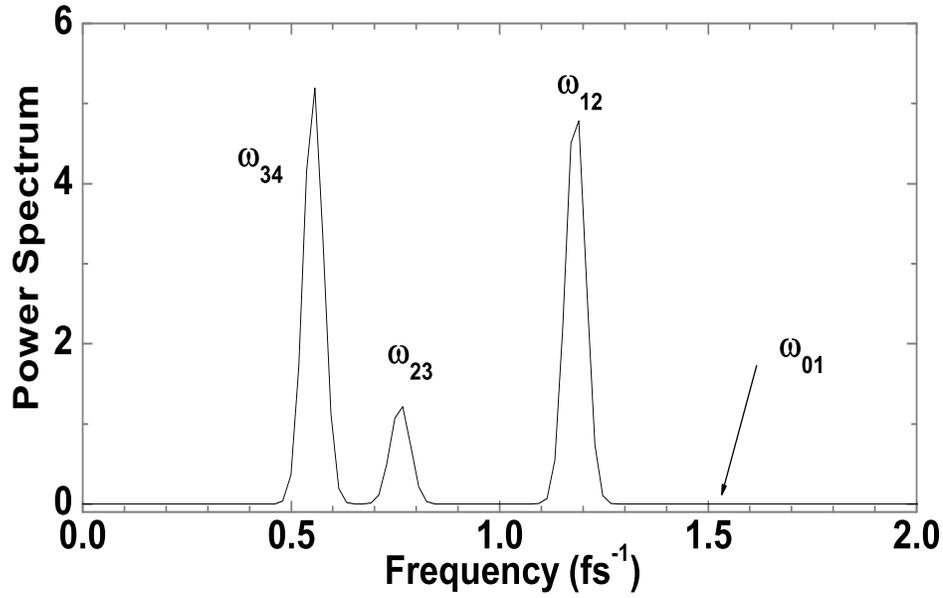}%
\caption{The optimal field power spectra of model 3. Here noise is only in the
amplitude of the $0\leftrightarrow1$ transition with $\Gamma_{A}=0.20$. The
optimal field cooperates with the noise by eliminating the $\omega_{01}$
frequency and exclusively relying on the noise to drive that step. The arrow
indicates where that frequency lies.}%
\end{center}
\end{figure}
\begin{figure}
[ptb]
\begin{center}
\includegraphics[
trim=0.000000in 0.000000in -1.701678in 0.000000in,
height=4.1174in,
width=6.1769in
]%
{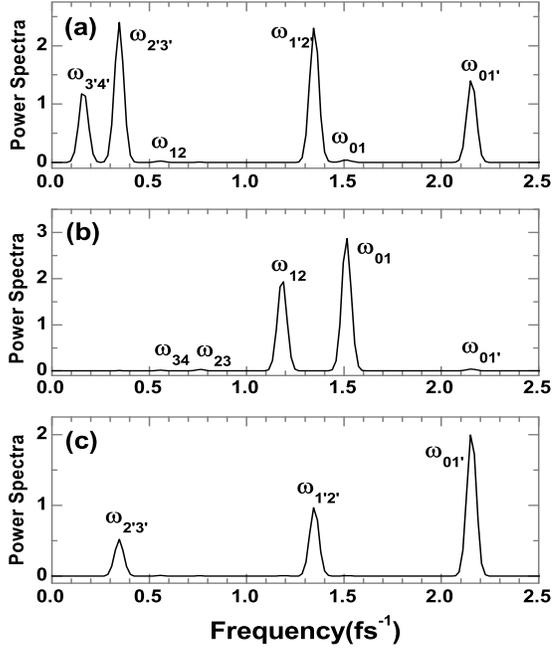}%
\caption{Power spectra of the control fields for model 4 (see panel (b) of
Figure 1). The frequencies along the right path in Figure 1 are denoted with a
prime to distinguish them from the left path. (a) No noise, (b) noise in the
left path, (c) noise in the right path. In these low target yield cases the
optimal field cooperates with the noise and follows the associated path.}%
\end{center}
\end{figure}


\bigskip

\begin{thebibliography}{99999999999999999999999999999999999999999999999999999999999999999999999999999999999999999999999999}
\expandafter\ifx\csname natexlab\endcsname\relax


\fi
\expandafter\ifx\csname bibnamefont\endcsname\relax


\fi
\expandafter\ifx\csname bibfnamefont\endcsname\relax


\fi
\expandafter\ifx\csname citenamefont\endcsname\relax


\fi
\expandafter\ifx\csname url\endcsname\relax


\fi
\expandafter\ifx\csname urlprefix\endcsname\relax


\fi
\providecommand{\bibinfo}[2]{#2} \providecommand{\eprint}[2][]{\url{#2}}

\bibitem[Rice and Zhao(2000)]{Rice00}%
\bibinfo{author}{\bibfnamefont{S.~A.} \bibnamefont{Rice}} and
\bibinfo{author}{\bibfnamefont{M.}~\bibnamefont{Zhao}},
\emph{\bibinfo{title}{Optical Control of Molecular Dynamics}}
(\bibinfo{publisher}{Wiley, New York}, \bibinfo{year}{2000}).

\bibitem[Rabitz(2003)]{Rabitz0364}%
\bibinfo{author}{\bibfnamefont{H.}~\bibnamefont{Rabitz}},
\bibinfo{journal}{Theor. Chem. Acc.} \textbf{\bibinfo{volume}{109}},
\bibinfo{pages}{64} (\bibinfo{year}{2003}).

\bibitem[Walmsley and Rabitz(2003)]{Walmsley0343}%
\bibinfo{author}{\bibfnamefont{I.}~\bibnamefont{Walmsley}} and
\bibinfo{author}{\bibfnamefont{H.}~\bibnamefont{Rabitz}},
\bibinfo{journal}{Phys. Today} \textbf{\bibinfo{volume}{56}},
\bibinfo{pages}{43} (\bibinfo{year}{2003}).

\bibitem[Brixner et~al.(2001{a})Brixner, Damrauer, and Gerber]{Brixner011}%
\bibinfo{author}{\bibfnamefont{T.}~\bibnamefont{Brixner}},
\bibinfo{author}{\bibfnamefont{N.~H.} \bibnamefont{Damrauer}}, and
\bibinfo{author}{\bibfnamefont{G.}~\bibnamefont{Gerber}}, in
\emph{\bibinfo{booktitle}{Advances in Atomic, Molecular, and Optical
Physics}}, edited by
\bibinfo{editor}{\bibfnamefont{B.}~\bibnamefont{Bederson}} and
\bibinfo{editor}{\bibfnamefont{H.}~\bibnamefont{Walther}}
(\bibinfo{publisher}{Academic}, \bibinfo{address}{San Diego, CA},
\bibinfo{year}{2001}{\natexlab{a}}), vol.~\bibinfo{volume}{46}, pp. \bibinfo{pages}{1--54}.

\bibitem[Zhu and Rabitz(2003)]{Wusheng036751}%
\bibinfo{author}{\bibfnamefont{W.}~\bibnamefont{Zhu}} and
\bibinfo{author}{\bibfnamefont{H.}~\bibnamefont{Rabitz}},
\bibinfo{journal}{J. Chem. Phys.} \textbf{\bibinfo{volume}{118}},
\bibinfo{pages}{6751} (\bibinfo{year}{2003}).

\bibitem[Geremia et~al.(2000)Geremia, Zhu, and Rabitz]{JM0010841}%
\bibinfo{author}{\bibfnamefont{J.~M.} \bibnamefont{Geremia}},
\bibinfo{author}{\bibfnamefont{W.}~\bibnamefont{Zhu}}, and
\bibinfo{author}{\bibfnamefont{H.}~\bibnamefont{Rabitz}},
\bibinfo{journal}{J. Chem. Phys.} \textbf{\bibinfo{volume}{113}},
\bibinfo{pages}{10841} (\bibinfo{year}{2000}).

\bibitem[Rabitz(2002)]{Rabitz0263405}%
\bibinfo{author}{\bibfnamefont{H.}~\bibnamefont{Rabitz}},
\bibinfo{journal}{Phys. Rev. A} \textbf{\bibinfo{volume}{66}},
\bibinfo{pages}{63405} (\bibinfo{year}{2002}).

\bibitem[Walser et~al.(1992)Walser, Ritsch, Zoller, and Cooper]{Walser92468}%
\bibinfo{author}{\bibfnamefont{R.}~\bibnamefont{Walser}},
\bibinfo{author}{\bibfnamefont{H.}~\bibnamefont{Ritsch}},
\bibinfo{author}{\bibfnamefont{P.}~\bibnamefont{Zoller}}, and
\bibinfo{author}{\bibfnamefont{J.}~\bibnamefont{Cooper}},
\bibinfo{journal}{Phys. Rev. A} \textbf{\bibinfo{volume}{45}},
\bibinfo{pages}{468} (\bibinfo{year}{1992}).

\bibitem[Sola and Rabitz(2004)]{Ignacio049009}%
\bibinfo{author}{\bibfnamefont{I.~R.} \bibnamefont{Sola}} and
\bibinfo{author}{\bibfnamefont{H.}~\bibnamefont{Rabitz}},
\bibinfo{journal}{J. Chem. Phys.} \textbf{\bibinfo{volume}{120}},
\bibinfo{pages}{9009} (\bibinfo{year}{2004}).

\bibitem[Akramine et~al.(1998)Akramine, Makhoute, Zitane, and Tij]%
{Akramine984892}\bibinfo{author}{\bibfnamefont{O.~E.} \bibnamefont{Akramine}},
\bibinfo{author}{\bibfnamefont{A.}~\bibnamefont{Makhoute}},
\bibinfo{author}{\bibfnamefont{M.}~\bibnamefont{Zitane}}, and
\bibinfo{author}{\bibfnamefont{M.}~\bibnamefont{Tij}},
\bibinfo{journal}{Phys. Rev. A} \textbf{\bibinfo{volume}{58}},
\bibinfo{pages}{4892} (\bibinfo{year}{1998}).

\bibitem[Akramine and Makhoute(1998)]{Akramine984349}%
\bibinfo{author}{\bibfnamefont{O.~E.} \bibnamefont{Akramine}} and
\bibinfo{author}{\bibfnamefont{A.}~\bibnamefont{Makhoute}},
\bibinfo{journal}{J. Phys. B} \textbf{\bibinfo{volume}{31}},
\bibinfo{pages}{4349} (\bibinfo{year}{1998}).

\bibitem[Judson and Rabitz(1992)]{Judson921500}%
\bibinfo{author}{\bibfnamefont{R.~S.} \bibnamefont{Judson}} and
\bibinfo{author}{\bibfnamefont{H.}~\bibnamefont{Rabitz}},
\bibinfo{journal}{Phys. Rev. Lett} \textbf{\bibinfo{volume}{68}},
\bibinfo{pages}{1500} (\bibinfo{year}{1992}).

\bibitem[Gammaitoni et~al.(1998)Gammaitoni, H\"{a}nggi, Jung, and
Marchesoni]{Gammaitoni98223}%
\bibinfo{author}{\bibfnamefont{L.}~\bibnamefont{Gammaitoni}},
\bibinfo{author}{\bibfnamefont{P.}~\bibnamefont{H\"{a}nggi}},
\bibinfo{author}{\bibfnamefont{P.}~\bibnamefont{Jung}}, and
\bibinfo{author}{\bibfnamefont{F.}~\bibnamefont{Marchesoni}},
\bibinfo{journal}{Rev. Mod. Phys.} \textbf{\bibinfo{volume}{70}},
\bibinfo{pages}{223} (\bibinfo{year}{1998}).

\bibitem[Zeidler et~al.(2001)Zeidler, Frey, Kompa, and Motzkus]%
{Zeidler0123420}\bibinfo{author}{\bibfnamefont{D.}~\bibnamefont{Zeidler}},
\bibinfo{author}{\bibfnamefont{S.}~\bibnamefont{Frey}},
\bibinfo{author}{\bibfnamefont{K.-L.} \bibnamefont{Kompa}}, and
\bibinfo{author}{\bibfnamefont{M.}~\bibnamefont{Motzkus}},
\bibinfo{journal}{Phys. Rev. A} \textbf{\bibinfo{volume}{64}},
\bibinfo{pages}{23420} (\bibinfo{year}{2001}).

\bibitem[Kunde et~al.(2000)Kunde, Baumann, Arlt, Morier-Genoud, Siegner, and
Keller]{Kunde00924}\bibinfo{author}{\bibfnamefont{J.}~\bibnamefont{Kunde}},
\bibinfo{author}{\bibfnamefont{B.}~\bibnamefont{Baumann}},
\bibinfo{author}{\bibfnamefont{S.}~\bibnamefont{Arlt}},
\bibinfo{author}{\bibfnamefont{F.}~\bibnamefont{Morier-Genoud}},
\bibinfo{author}{\bibfnamefont{U.}~\bibnamefont{Siegner}}, and
\bibinfo{author}{\bibfnamefont{U.}~\bibnamefont{Keller}},
\bibinfo{journal}{Appl. Phys. Lett.} \textbf{\bibinfo{volume}{77}},
\bibinfo{pages}{924} (\bibinfo{year}{2000}).

\bibitem[Brixner et~al.(2001{b})Brixner, Damrauer, Niklaus, and Gerber]%
{Brixner0157}\bibinfo{author}{\bibfnamefont{T.}~\bibnamefont{Brixner}},
\bibinfo{author}{\bibfnamefont{N.~H.} \bibnamefont{Damrauer}},
\bibinfo{author}{\bibfnamefont{P.}~\bibnamefont{Niklaus}}, and
\bibinfo{author}{\bibfnamefont{G.}~\bibnamefont{Gerber}},
\bibinfo{journal}{Nature (London)} \textbf{\bibinfo{volume}{414}},
\bibinfo{pages}{57} (\bibinfo{year}{2001}{\natexlab{b}}).

\bibitem[Levis et~al.(2001)Levis, Menkir, and Rabitz]{Levis01709}%
\bibinfo{author}{\bibfnamefont{R.~J.} \bibnamefont{Levis}},
\bibinfo{author}{\bibfnamefont{G.~M.} \bibnamefont{Menkir}}, and
\bibinfo{author}{\bibfnamefont{H.}~\bibnamefont{Rabitz}},
\bibinfo{journal}{Science} \textbf{\bibinfo{volume}{292}},
\bibinfo{pages}{709} (\bibinfo{year}{2001}).

\bibitem[Daniel et~al.(2003)Daniel, Full, Gonz\'{a}lez, Lupulescu, Manz,
Merli, \v{S}. Vajda, and W\"{o}ste]{Daniel03536}%
\bibinfo{author}{\bibfnamefont{C.}~\bibnamefont{Daniel}},
\bibinfo{author}{\bibfnamefont{J.}~\bibnamefont{Full}},
\bibinfo{author}{\bibfnamefont{L.}~\bibnamefont{Gonz\'{a}lez}},
\bibinfo{author}{\bibfnamefont{C.}~\bibnamefont{Lupulescu}},
\bibinfo{author}{\bibfnamefont{J.}~\bibnamefont{Manz}},
\bibinfo{author}{\bibfnamefont{A.}~\bibnamefont{Merli}},
\bibinfo{author}{\bibnamefont{\v{S}. Vajda}}, and
\bibinfo{author}{\bibfnamefont{L.}~\bibnamefont{W\"{o}ste}},
\bibinfo{journal}{Science} \textbf{\bibinfo{volume}{299}},
\bibinfo{pages}{536} (\bibinfo{year}{2003}).

\bibitem[Assion et~al.(1998)Assion, Baumert, Bergt, Brixner, Kiefer, Seyfried,
Strehle, and Gerber]{Gerber98919}%
\bibinfo{author}{\bibfnamefont{A.}~\bibnamefont{Assion}},
\bibinfo{author}{\bibfnamefont{T.}~\bibnamefont{Baumert}},
\bibinfo{author}{\bibfnamefont{M.}~\bibnamefont{Bergt}},
\bibinfo{author}{\bibfnamefont{T.}~\bibnamefont{Brixner}},
\bibinfo{author}{\bibfnamefont{B.}~\bibnamefont{Kiefer}},
\bibinfo{author}{\bibfnamefont{V.}~\bibnamefont{Seyfried}},
\bibinfo{author}{\bibfnamefont{M.}~\bibnamefont{Strehle}}, and
\bibinfo{author}{\bibfnamefont{G.}~\bibnamefont{Gerber}},
\bibinfo{journal}{Science} \textbf{\bibinfo{volume}{282}},
\bibinfo{pages}{919} (\bibinfo{year}{1998}).

\bibitem[Grossmann et~al.(1993)Grossmann, Dittrich, Jung, and H\"{a}nggi]%
{Grossmann93229}\bibinfo{author}{\bibfnamefont{F.}~\bibnamefont{Grossmann}},
\bibinfo{author}{\bibfnamefont{T.}~\bibnamefont{Dittrich}},
\bibinfo{author}{\bibfnamefont{P.}~\bibnamefont{Jung}}, and
\bibinfo{author}{\bibfnamefont{P.}~\bibnamefont{H\"{a}nggi}},
\bibinfo{journal}{J. Stat. Phys.} \textbf{\bibinfo{volume}{70}},
\bibinfo{pages}{229} (\bibinfo{year}{1993}).

\bibitem[Blanchard et~al.(1994)Blanchard, Bolz, Cini, Deangelis, and
Serva]{Blanchard94749}%
\bibinfo{author}{\bibfnamefont{P.}~\bibnamefont{Blanchard}},
\bibinfo{author}{\bibfnamefont{G.}~\bibnamefont{Bolz}},
\bibinfo{author}{\bibfnamefont{M.}~\bibnamefont{Cini}},
\bibinfo{author}{\bibfnamefont{G.~F.} \bibnamefont{Deangelis}}, and
\bibinfo{author}{\bibfnamefont{M.}~\bibnamefont{Serva}},
\bibinfo{journal}{J. Stat. Phys.} \textbf{\bibinfo{volume}{75}},
\bibinfo{pages}{749} (\bibinfo{year}{1994}).

\bibitem[Shao et~al.(1998)Shao, Zerbe, and H\"{a}nggi]{Shao9781}%
\bibinfo{author}{\bibfnamefont{J.~S.} \bibnamefont{Shao}},
\bibinfo{author}{\bibfnamefont{C.}~\bibnamefont{Zerbe}}, and
\bibinfo{author}{\bibfnamefont{P.}~\bibnamefont{H\"{a}nggi}},
\bibinfo{journal}{Chem. Phys.} \textbf{\bibinfo{volume}{235}},
\bibinfo{pages}{81} (\bibinfo{year}{1998}).

\bibitem[Klappauf et~al.(1998)Klappauf, Oskay, Steck, and Raizen]%
{Klappauf981203}\bibinfo{author}{\bibfnamefont{B.~G.} \bibnamefont{Klappauf}},
\bibinfo{author}{\bibfnamefont{W.~H.} \bibnamefont{Oskay}},
\bibinfo{author}{\bibfnamefont{D.~A.} \bibnamefont{Steck}}, and
\bibinfo{author}{\bibfnamefont{M.~G.} \bibnamefont{Raizen}},
\bibinfo{journal}{Phys. Rev. Lett.} \textbf{\bibinfo{volume}{81}},
\bibinfo{pages}{1203} (\bibinfo{year}{1998}).

\bibitem[Gross et~al.(1992)Gross, Neuhauser, and Rabitz]{Gross924557}%
\bibinfo{author}{\bibfnamefont{P.}~\bibnamefont{Gross}},
\bibinfo{author}{\bibfnamefont{D.}~\bibnamefont{Neuhauser}}, and
\bibinfo{author}{\bibfnamefont{H.}~\bibnamefont{Rabitz}},
\bibinfo{journal}{J. Chem. Phys.} \textbf{\bibinfo{volume}{98}},
\bibinfo{pages}{4557} (\bibinfo{year}{1992}).

\bibitem[T\'{o}th et~al.(1994)T\'{o}th, L\H{o}rincz, and Rabitz]%
{Toth943715}\bibinfo{author}{\bibfnamefont{G.~J.} \bibnamefont{T\'{o}th}},
\bibinfo{author}{\bibfnamefont{A.}~\bibnamefont{L\H{o}rincz}}, and
\bibinfo{author}{\bibfnamefont{H.}~\bibnamefont{Rabitz}},
\bibinfo{journal}{J. Chem. Phys.} \textbf{\bibinfo{volume}{101}},
\bibinfo{pages}{3715} (\bibinfo{year}{1994}).

\bibitem[Shuang et~al.(2000)Shuang, Yang, Zhang, and Yan]{Shuang007192}%
\bibinfo{author}{\bibfnamefont{F.}~\bibnamefont{Shuang}},
\bibinfo{author}{\bibfnamefont{C.}~\bibnamefont{Yang}},
\bibinfo{author}{\bibfnamefont{H.}~\bibnamefont{Zhang}}, and
\bibinfo{author}{\bibfnamefont{Y.}~\bibnamefont{Yan}},
\bibinfo{journal}{Phys. Rev. E} \textbf{\bibinfo{volume}{61}},
\bibinfo{pages}{7192} (\bibinfo{year}{2000}).

\bibitem[Vugmeister and Rabitz(1997)]{Vugmeister972522}%
\bibinfo{author}{\bibfnamefont{B.~E.} \bibnamefont{Vugmeister}} and
\bibinfo{author}{\bibfnamefont{H.}~\bibnamefont{Rabitz}},
\bibinfo{journal}{Phys. Rev. E} \textbf{\bibinfo{volume}{55}},
\bibinfo{pages}{2522} (\bibinfo{year}{1997}).

\bibitem[Smelyanskiy and Dykman(1997)]{Mark972516}%
\bibinfo{author}{\bibfnamefont{V.~N.} \bibnamefont{Smelyanskiy}} and
\bibinfo{author}{\bibfnamefont{M.~I.}
\bibnamefont{Dykman}}, \bibinfo{journal}{Phys. Rev. E}
\textbf{\bibinfo{volume}{55}}, \bibinfo{pages}{2516} (\bibinfo{year}{1997}).

\bibitem[Pechukas and H\"{a}nggi(1994)]{Pechukas942772}%
\bibinfo{author}{\bibfnamefont{P.}~\bibnamefont{Pechukas}} and
\bibinfo{author}{\bibfnamefont{P.}~\bibnamefont{H\"{a}nggi}},
\bibinfo{journal}{Phys. Rev. Lett.} \textbf{\bibinfo{volume}{73}},
\bibinfo{pages}{2772} (\bibinfo{year}{1994}).

\bibitem[Mantegna and Spagnolo(1996)]{Mantegna96563}%
\bibinfo{author}{\bibfnamefont{R.~N.} \bibnamefont{Mantegna}} and
\bibinfo{author}{\bibfnamefont{B.}~\bibnamefont{Spagnolo}},
\bibinfo{journal}{Phys. Rev. Lett.} \textbf{\bibinfo{volume}{76}},
\bibinfo{pages}{563} (\bibinfo{year}{1996}).

\bibitem[Mantegna and Spagnolo(2000)]{Mantegna003025}%
\bibinfo{author}{\bibfnamefont{R.~N.} \bibnamefont{Mantegna}} and
\bibinfo{author}{\bibfnamefont{B.}~\bibnamefont{Spagnolo}},
\bibinfo{journal}{Phys. Rev. Lett.} \textbf{\bibinfo{volume}{84}},
\bibinfo{pages}{3025} (\bibinfo{year}{2000}).

\bibitem[Shuang et~al.()Shuang, Dykman, and Rabitz]{Shuang2004_2}%
\bibinfo{author}{\bibfnamefont{F.}~\bibnamefont{Shuang}},
\bibinfo{author}{\bibfnamefont{M.~I.} \bibnamefont{Dykman}}, and
\bibinfo{author}{\bibfnamefont{H.}~\bibnamefont{Rabitz}}, \bibinfo{note}{unpublished}.

\bibitem[Goldberg(1989)]{Goldberg97}%
\bibinfo{author}{\bibfnamefont{D.~E.} \bibnamefont{Goldberg}},
\emph{\bibinfo{title}{Genetic Algorithms in Search, Optimization, and Machine
Learning}} (\bibinfo{publisher}{Addison-Wesley}, \bibinfo{address}{Reading,
MA}, \bibinfo{year}{1989}).

\bibitem[Aizawa and Wah(1993)]{Aizawa9348}%
\bibinfo{author}{\bibfnamefont{A.~N.} \bibnamefont{Aizawa}} and
\bibinfo{author}{\bibfnamefont{B.~W.} \bibnamefont{Wah}}, in
\emph{\bibinfo{booktitle}{Proceedings of the Fifth international conference
on Genetic Algorithms, Urbana-Champaign, IL, USA, June 1993}}, edited by
\bibinfo{editor}{\bibfnamefont{S.}~\bibnamefont{Forrest}}
(\bibinfo{publisher}{Morgan Kaufmann}, \bibinfo{address}{San Mateo, CA},
\bibinfo{year}{1993}), pp. \bibinfo{pages}{48--55}.

\bibitem[Miller(1997)]{Miller97}%
\bibinfo{author}{\bibfnamefont{B.~L.} \bibnamefont{Miller}}, Ph.D. thesis,
\bibinfo{school}{University of Illinois at Urbana-Champaign},
\bibinfo{address}{Urbana} (\bibinfo{year}{1997}).

\bibitem[Seijas et~al.(2002)Seijas, Morat\'{o}, and Sanz-Gonz\'{a}lez]%
{Seijas02617}\bibinfo{author}{\bibfnamefont{J.}~\bibnamefont{Seijas}},
\bibinfo{author}{\bibfnamefont{C.}~\bibnamefont{Morat\'{o}}}, and
\bibinfo{author}{\bibfnamefont{J.~L.}
\bibnamefont{Sanz-Gonz\'{a}lez}}, in
\emph{\bibinfo{booktitle}{Proceedings of
the Fourth International Conference on Parallel Processing and Applied
Mathematics}}, edited by \bibinfo{editor}{\bibnamefont{V.R.Wyrzykowski}},
\bibinfo{editor}{\bibfnamefont{J.}~\bibnamefont{Dongarra}},
\bibinfo{editor}{\bibnamefont{M.Paprzycki}}, and
\bibinfo{editor}{\bibnamefont{J.Wasniewski}} (\bibinfo{publisher}{Springer},
\bibinfo{address}{London, UK}, \bibinfo{year}{2002}), pp. \bibinfo{pages}{617--625}.

\bibitem[Yip et~al.(2003)Yip, Mazziotti, and Rabitz]{Yip038168}%
\bibinfo{author}{\bibfnamefont{F.}~\bibnamefont{Yip}},
\bibinfo{author}{\bibfnamefont{D.}~\bibnamefont{Mazziotti}}, and
\bibinfo{author}{\bibfnamefont{H.}~\bibnamefont{Rabitz}},
\bibinfo{journal}{J. Chem. Phys.} \textbf{\bibinfo{volume}{118}},
\bibinfo{pages}{8168} (\bibinfo{year}{2003}).

\bibitem[Rana et~al.(1996)Rana, Whitley, and Cogswell]{Rana96198}%
\bibinfo{author}{\bibfnamefont{S.}~\bibnamefont{Rana}},
\bibinfo{author}{\bibfnamefont{L.~D.} \bibnamefont{Whitley}}, and
\bibinfo{author}{\bibfnamefont{R.}~\bibnamefont{Cogswell}}, in
\emph{\bibinfo{booktitle}{Proceedings of the Fourth International
Conference on Parallel Problem Solving from Nature}}, edited by
\bibinfo{editor}{\bibfnamefont{H.}~\bibnamefont{Voigt}},
\bibinfo{editor}{\bibfnamefont{W.}~\bibnamefont{Ebeling}},
\bibinfo{editor}{\bibfnamefont{I.}~\bibnamefont{Rechenberg}}, and
\bibinfo{editor}{\bibfnamefont{H.-P.}
\bibnamefont{Schwefel}} (\bibinfo{publisher}{Springer},
\bibinfo{address}{Berlin}, \bibinfo{year}{1996}), pp. \bibinfo{pages}{198--207}.
\end{thebibliography}
\end{document}